\documentclass[twocolumn]{revtex4}
\usepackage{amsfonts}
\usepackage{amsmath}
\usepackage{amssymb,epsf}
\usepackage{latexsym}
\usepackage{graphicx,epsfig}
\usepackage{amssymb}
\usepackage{float}
\usepackage{subfigure}
\usepackage{epstopdf}
\usepackage[colorlinks=true,citecolor=blue,linkcolor=blue,urlcolor=black]{hyperref}
\usepackage{dcolumn}
\usepackage{psfrag}
\usepackage{wrapfig}
\usepackage{makeidx}
\usepackage{epsf}
\usepackage{color}

\graphicspath{{Images/}}
\begin{document}

\title{A note on cosmological features of modified Newtonian potentials}
\author{M. Kord Zangeneh}
\email{mkzangeneh@scu.ac.ir}
\affiliation{Physics Department, Faculty of Science, Shahid Chamran University of Ahvaz,
Ahvaz 61357-43135, Iran}
\author{H. Moradpour}
\email{h.moradpour@riaam.ac.ir}
\affiliation{Research Institute for Astronomy and Astrophysics of Maragha (RIAAM),
Maragha 55134-441, Iran}
\author{N. Sadeghnezhad}
\affiliation{Research Institute for Astronomy and Astrophysics of Maragha (RIAAM),
Maragha 55134-441, Iran}

\begin{abstract}
Considering some modified Newtonian potentials and the Hubble law
in writing the total energy of a test mass located at the edge of
a flat Friedmann-Robertson-Walker universe, we obtain several
modified Friedmann equations. Interestingly enough, our study
shows that the employed potentials, while some of them have some
successes in modelling the spiral galaxies rotation curves, may
also address an accelerated universe. This fact indicates that
dark energy and dark matter may have some common origins and
aspects.
\end{abstract}

\maketitle

\section{Introduction\label{Intr}}

One of the interesting problems in cosmology is seeking for
modified forms of the Newtonian potential which can cover the
results of the general relativity (GR) in the Newtonian framework
\cite{nc6,ncprd,fes,yc1,yc2,yc4,yc5,yc6,yc7,yc8,yc9,roos,ybh,ej,yp,m1,m2,m3,m4,m5,m6,m7,m8,ref1}.
A famous example, which originally returns to Newton, is
\cite{m1,m2}

\begin{equation}
V\left( r\right) =\left( A+\frac{B}{r}\right) V_{\mathrm{N}}\left( r\right) ,%
\text{ Type A}  \label{e0}
\end{equation}

\noindent where $V_{\mathrm{N}}\left( r\right) =-GmM/r$ denotes
the Newtonian potential and $A$ and $B$ are unknown parameters
found by either fitting with the observations
\cite{m1,m2,m3,m4,m5,m6,m7,m8} or using other parts of physics
\cite{m9}. In cosmological setup, without working in the GR
framework, one can still get the Friedmann equations by using the
Hubble law and the Newtonian potential in order to write the total
energy of a test particle located at the edge of the
Friedmann-Robertson-Walker (FRW) universe
\cite{nm,nc1,nc2,nc20,nc6,nc4,nc5,ncprd,roos}. Nevertheless, the
Newtonian potential cannot provide suitable description for some
questions
such as the backreaction, dark matter and dark energy problems \cite%
{ncprd,roos,nc6} and one needs alternative theories of Newtonian
gravity to avoid these difficulties \cite{ncprd}.

In fact, due to pointed out problems, modified versions of the
Newtonian potential such as the Yukawa potential \cite{yp} have
been employed. These potentials lead to interesting results in
describing gravitational phenomena
\cite{ncprd,fes,yc1,yc2,yc4,yc5,yc6,yc7,yc8,yc9,ybh,ej}. Amonog
different kinds of modified potentials, the below ones have
attracted more attentions
in studying the gravitational systems \cite{fes,yc1,yp,ybh,ej,m1}%
\begin{equation}
V\left( r\right) =\left\{
\begin{array}{ll}
V_{\mathrm{N}}\left( r\right) e^{-\alpha r}, & \text{Type B} \\
&  \\
\left( 1+\beta e^{-\alpha r}\right) V_{\mathrm{N}}\left( r\right) & \text{%
Type C}%
\end{array}%
\right.  \label{cp}
\end{equation}

\noindent where $\alpha $ and $\beta $ are some constants. The
possible ranges for the values of $\alpha $ and $\beta $ depend on
the system. These potentials can support some kinds of black holes
\cite{ybh}. It is also useful to mention that the potentials such
as Type C can be obtained either
by looking for the Newtonian limits of some modified GR theories \cite%
{yc31,yc32,yc3} or taking the probable non-local features of the
Newtonian gravity into account \cite{ynl}. Further corrections to
the Newtonian
potential such as the logarithmic modifications can be found in Refs.~\cite%
{ynl,log1,log2,log3,r1}. The latter correction can successfully
describe the spiral galaxies rotation curves. It is also useful to
mention here that the gravitational wave (GW) astronomy is a very
powerful tool to find out the more comprehensive gravitational
theory \cite{corda1,corda2}.

The ability of the mentioned potentials to describe the current
accelerated universe has not been studied yet. So, the question
whether the modifications to the Newtonian potential can provide a
classical description for the dark energy remains unrevealed. In
addition to this query, it is also interesting to study the
cosmological consequences of the above potentials to figure out if
they can provide acceptable descriptions for
various gravitational systems \cite%
{ncprd,yc1,yc2,yc4,yc5,yc6,yc7,yc8,yc9,fes,ybh,ej,m1,m2,m3,m4,m5,m6,m7,m8,m9}%
.

In the next section, we introduce and study three sets of modified
Friedmann
equations corresponding to the above introduced potentials. Next, in Sec.~%
\ref{logsec}, after addressing a Hook correction to the Newtonian
potential and investigating some of its properties, the
cosmological consequences of a logarithmic modified Newtonian
potential will be studied. The last section is devoted to the
summary and conclusion. Throughout this paper, dot denotes the
derivative with respect to time and we set $c=\hbar =1$.

\section{Modified Newtonian models and dark energy}

Consider an expanding box with radius $R$, filled by a fluid with
energy density $\rho $, while there is a test mass $m$ on its edge
\cite{roos}. In this manner, the Hubble law leads to $v=HR$ ($H$
is the Hubble parameter) for the velocity of the test particle
\cite{roos}. This situation is a non-relativistic counterpart of
considering a FRW universe with scale factor $a(t)$ and Hubble
parameter $H\equiv \frac{\dot{a}}{a}$, enclosed by its apparent
horizon, and filled by a fluid with energy density $\rho $ while
the test mass $m$ is located on the apparent horizon \cite{roos}.
For this setup, $r_{h}=a(t)r_{c}\equiv R$ is the radius of the
apparent horizon and the system aerial volume is $V=\frac{4\pi
}{3}r_{h}^{3}$ where $r_{c}$
denotes the co-moving radius of the apparent horizon \cite%
{ijmpd,roos,jhep,jhep1}. Since WMAP data indicates a flat universe ($r_{h}=%
\frac{1}{H}$) \cite{roos}, we only consider the flat universe for
which the aerial volume is equal to the real volume
\cite{jhep,jhep1}.

For a test particle with mass $m$ located at the apparent horizon
(or equally the edge of the expanding box), using the Hubble law,
one can write
the relation between the particle velocity ($v$), the particle distance ($%
R=r_{h}$) and the Hubble parameter as $v=Hr_{h}$ and thus
\cite{roos}

\begin{equation}
E=\frac{1}{2}mH^{2}r_{h}^{2}+V(r_{h}),  \label{e1}
\end{equation}

\noindent is the total energy of the test particle. Using the
potentials introduced in (\ref{e0}) and (\ref{cp}) and the total
mass

\begin{equation}
M=\int \rho dV,  \label{tm}
\end{equation}

\noindent Eq.~(\ref{e1}) leads to%
\begin{equation}
H^{2}=\left\{
\begin{array}{ll}
-\frac{K}{a^{2}}+\frac{8\pi G}{3}\rho \left(
A+\frac{B}{r_{h}}\right) &
\text{Type A} \\
&  \\
-\frac{K}{a^{2}}+\frac{8\pi G}{3}\rho e^{-\alpha r_{h}}, & \text{Type B} \\
&  \\
-\frac{K}{a^{2}}+\frac{8\pi G}{3}\rho \left( 1+\beta e^{-\alpha
r_{h}}\right)
& \text{Type C}%
\end{array}%
\right.  \label{e2}
\end{equation}%
where $K\equiv -2E/mr_{c}^{2}$ \cite{nm,nc1,nc2,nc20,nc6,nc4,nc5,ncprd,roos}%
. One could easily confirm that in the absence of the correction
terms, relations in (\ref{e2}) recovers the Friedmann equation (at
least mathematically) only if the role of curvature constant of
the FRW universe is attributed to $K$
\cite{nm,nc1,nc2,nc20,nc6,nc4,nc5,ncprd,roos}. Of course, for type
A, we should fix $A=1$. Since we intend to consider the
case similar to flat FRW universe, we have to set $K=0$ and $r_{h}=1/H$, so%
\begin{equation}
H^{2}=\left\{
\begin{array}{ll}
\frac{8\pi G}{3}\rho \left( 1+BH\right) & \text{Type A} \\
&  \\
-\frac{8\pi G}{3}\rho e^{-\frac{\alpha }{H}}, & \text{Type B} \\
&  \\
-\frac{8\pi G}{3}\rho \left( 1+\beta e^{-\frac{\alpha }{H}}\right) & \text{%
Type C}%
\end{array}%
\right.  \label{ye1}
\end{equation}%
In the remaining of this section, we will study the ability of
these models to describe the dark energy effects.

\subsection{Type A}

By making the definition for density due to type A modification as%
\begin{equation}
\rho _{A}=\frac{3H^{2}}{8\pi G}\left( \frac{BH}{1+BH}\right) ,
\end{equation}
one could rewrite the first relation in (\ref{ye1}) as

\begin{equation}
3H^{2}=8\pi G\left( \rho +\rho _{A}\right) .
\end{equation}%
So, the density parameter is $\Omega _{A}\equiv 8\pi G\rho
_{A}/3H^{2}=BH/\left( 1+BH\right) $. Clearly, since $H\geq 0$
during the cosmic evolution, $\Omega _{A}$ is always positive only
if $B>0$. Therefore, density parameter $\Omega _{A}$ decreases as
$H$ decreases ($d\Omega _{A}/dH>0$). It means that $\rho _{A}$
cannot play the role of dark energy in the current universe.

\subsection{Type B}

We consider a background filled by a pressureless source
\cite{nc4} with energy density $\rho =\rho _{0}a^{-3}=\rho
_{0}(1+z)^{3}$, where $\rho _{0}$ is the current value of the dust
density \cite{roos}, $a$ is scale factor and $z=a^{-1}-1$ is
redshift. Now, using the second relation in~(\ref{ye1}), and by
defining the density parameter $\Omega _{B}=1-e^{\frac{\alpha
}{H}}$ (or equally $\rho _{B}=3H^{2}(1-e^{\frac{\alpha }{H}})/8\pi
G$), one finds

\begin{equation}
z(\Omega _{B})=\left( \frac{\gamma (1-\Omega _{B})}{\ln ^{2}(1-\Omega _{B})}%
\right) ^{1/3}-1,  \label{zo0}
\end{equation}

\noindent where $\gamma =3\alpha ^{2}/8\pi G\rho _{0}$ is an
unknown
parameter found by fitting the theory with observations. If the value of $%
\rho _{0}$ is known, finding possible values for $\gamma $ leads
to possible values for $\alpha $. It is also worthwhile mentioning
that since $\alpha ^{2}$ and $\rho _{0}$ are positive, Eq.
(\ref{zo0}) implies that $z\geq -1$
for $\Omega _{B}\leq 1$. In addition, by taking the second relation of (\ref%
{ye1}) into account, one can easily obtain

\begin{eqnarray}
&&3H^{2}=8\pi G\left( \rho +\rho _{B}\right) ,  \label{py1} \\
&&  \notag \\
&&3H^{2}+2\dot{H}=-8\pi Gp_{B},  \label{py12}
\end{eqnarray}

\noindent where

\begin{eqnarray}
&&p_{B}=-(\frac{\dot{\rho}_{B}}{3H}+\rho _{B})=  \notag  \label{py2} \\
&&-\left[ \rho _{B}+\frac{\dot{H}}{H^{2}}\left( \rho _{B}(2-\frac{\alpha }{H}%
)+\frac{3\alpha H}{8\pi G}\right) \right] ,
\end{eqnarray}

\noindent which clearly shows that $p_{B}\rightarrow -\rho _{B}$ when $\dot{H%
}\rightarrow 0$. In this manner, it is easy to see that $\Omega _{B}=1-e^{%
\frac{\alpha }{H}}=8\pi G\rho _{B}/3H^{2}$ is indeed the density
parameter corresponding to a fluid with energy density $\rho
_{B}$. In fact, by this way, we simulated the modification to
Newtonian potential as a hypothetical fluid with energy density
$\rho _{B}$ and pressure $p_{B}$ which has no interaction with the
pressureless energy source $\rho $. In addition, the
deceleration and total state parameters of type B model are also defined as $%
q_{B}=-1+\frac{1+z}{H(z)}\frac{dH(z)}{dz}$ and $\omega _{B}\equiv \frac{p_{B}%
}{\rho _{B}+\rho }=\frac{2}{3}(q_{B}-\frac{1}{2})$. Differentiating Eq. (\ref%
{py1}) with respect to $z$ and using the definitions for $q_{B}$,
$\rho
\left( z\right) $, $\rho _{B}$ and $\Omega _{B}$ one could obtain%
\begin{equation}
q_{B}=\frac{3(1+z)^{3}\Upsilon ^{2}}{\gamma (\Omega
_{B}-1)(\Upsilon -2)}-1.
\end{equation}%
where $\Upsilon =\ln (1-\Omega _{B})$.

\begin{figure}[t]
\centering
\includegraphics[width=.45\textwidth]{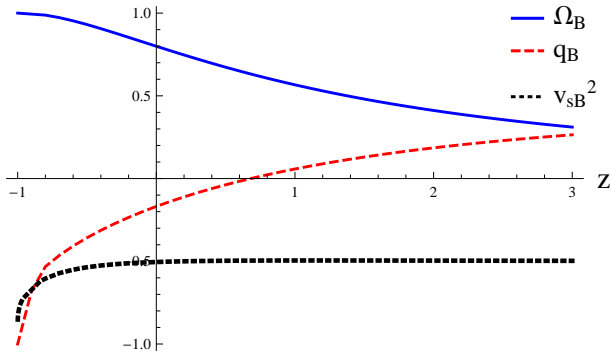}
\caption{The behaviors of $\Omega _{B}$ (solid line), the
deceleration parameter $q_{B}$ (dashed line) and squared speed of
sound $v_{sB}^{2}$ (dotted line) in term of $z$ with
$\protect\gamma =12\cdot 9$ for type B modified Newtonian model.}
\label{py}
\end{figure}

In Fig. (\ref{py}), $q_{B}$ and $\Omega _{B}$ have been plotted
for $\gamma =12\cdot 9$. For this value of $\gamma $, we have
$z_{t}\simeq 0\cdot 68$ where $z_{t}$ is transition redshift at
which the universe leaves the matter dominated era and enters an
accelerated era (or equally $q(z_{t})=0$). It is
remarkable to mention that, for $\gamma >0$, $q_{B}\rightarrow -1$ and $%
\Omega _{B}\rightarrow 1$ as $z\rightarrow -1$ and $q_{B}\rightarrow \frac{1%
}{2}$ and $\Omega _{B}\rightarrow 0$ for $z\gg 1$. Therefore,
bearing the mutual relation between $\omega _{B}$ and $q_{B}$ in
mind, one can easily find that $\omega _{B}(z\rightarrow
-1)\rightarrow -1$ and $\omega _{B}(z\gg 1)\rightarrow 0$ that is
a desired result. Moreover, our numerical
calculations show that whenever $9\cdot 5\leq \gamma \leq 23$, we have $%
0\cdot 5\leq z_{t}\leq 1$, $0\cdot 771\leq \Omega _{B}(z=0)\leq
0\cdot 847$ and $-0\cdot 1366\leq q_{B}(z=0)\leq -0\cdot 2261$.
For the current universe, observations indicate $q_{B}(z=0)\leq
-\frac{1}{2}$ \cite{roos}. This result is obtainable in type B
model if $\gamma >23$. However, in this case, $z_{t}>1$ and
$\Omega _{B}(z=0)>0\cdot 847$. In order to study the classical
stability of the obtained dark energy candidate, one should use
the squared speed of sound

\begin{eqnarray}
v_{sB}^{2} &=&\frac{dp_{B}}{d\rho _{B}}=\frac{dp_{B}/dH}{d\rho
_{B}/dH}
\notag \\
&=&\frac{2\Upsilon ^{-1}\left( \Upsilon ^{-1}-1\right) \left(
1-2\Upsilon ^{-1}\right) ^{-2}}{1-\Omega _{B}\left( 1-2\Upsilon
^{-1}\right) }.
\end{eqnarray}%
$v_{sB}^{2}$ is plotted in Fig.~(\ref{py}). It shows instability for all $%
z $ values specially at present time $z=0$. This is a common
behavior for many of the models for dark energy
\cite{stab,istab,istab1}.

\subsection{Type C}

\begin{figure*}[t]
\includegraphics[width=.46\textwidth]{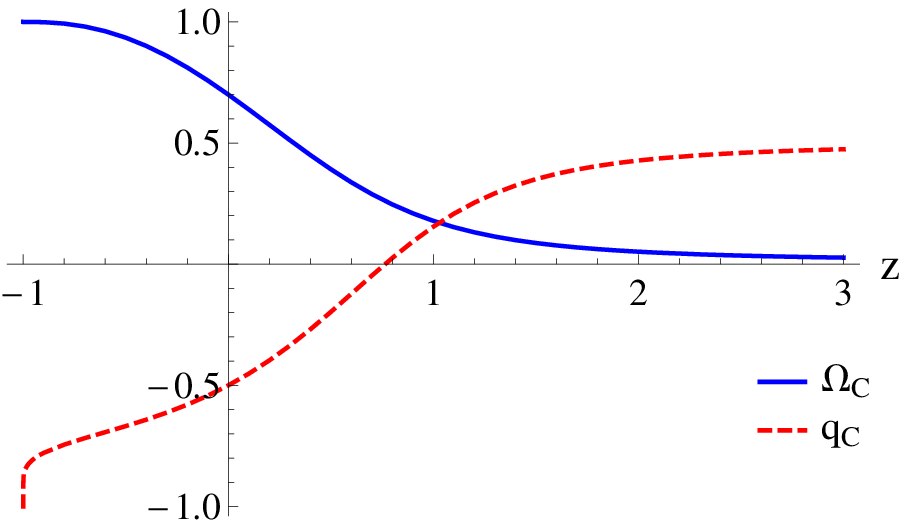}\label{pcy}
\includegraphics[width=.46\textwidth]{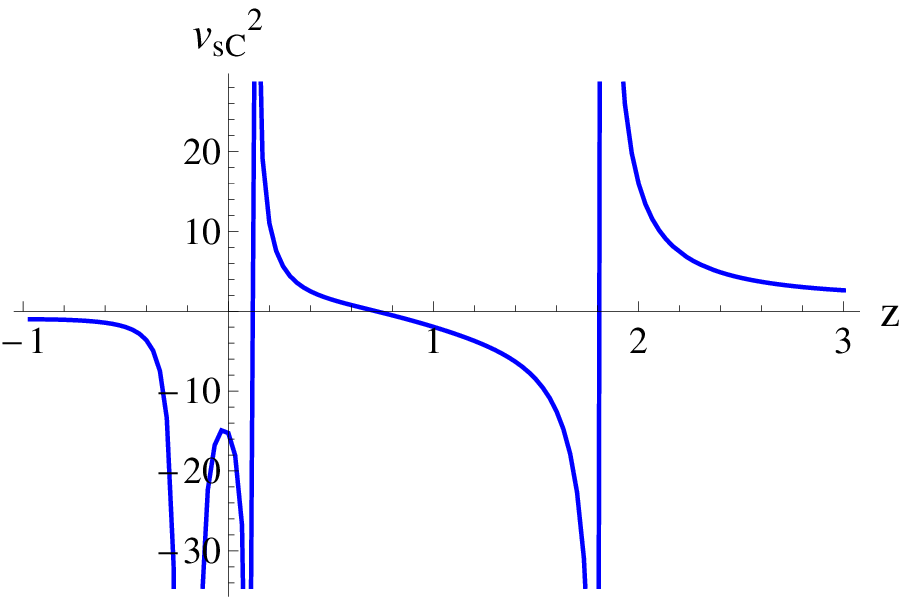}\label{stab1}
\caption{The behaviors of $\Omega_{C}$, $q_{C}$ and $v_{sC}^{2}$
in term of $z$ with $\protect\gamma =108\cdot 843537$ and
$\protect\beta =0\cdot 007696$} \label{fig1}
\end{figure*}

Now, we focus on the third model obtained by using the second potential in (%
\ref{cp}). The third relation in~(\ref{ye1}) can be rewritten as

\begin{equation}
3H^{2}=8\pi G\left( \rho +\rho _{C}\right) ,  \label{e4}
\end{equation}%
where $\rho _{C}=3\beta H^{2}e^{-\frac{\alpha }{H}}/8\pi G\left( 1+\beta e^{-%
\frac{\alpha }{H}}\right) $. Indeed, we stored the effects of
deviation from the Newtonian potential into $\rho _{C}$. Again, we
consider a dust source
\cite{nc4} satisfying ordinary energy-momentum conservation law ($\dot{\rho}%
_{C}=-3H\rho _{C}$). Thus, if an unknown pressure $p_{C}$ is
attributed to the hypothetical fluid with energy density $\rho
_{C}$, then we should have

\begin{equation}
\dot{\rho}_{C}+3H(\rho _{C}+p_{C})=0.  \label{emc1}
\end{equation}%
Now, combining the above relations with each other, one receives

\begin{eqnarray}
&&3H^{2}+2\dot{H}=-8\pi Gp_{C},  \label{e41} \\
&&  \notag \\
&&\dot{H}=-4\pi G(\rho +\rho _{C}+p_{C}),
\end{eqnarray}%
in which

\begin{eqnarray}
&&p_{C}=-(\frac{\dot{\rho}_{C}}{3H}+\rho _{C})=  \notag \\
&&-\frac{\rho _{C}}{3}\left[ \frac{\dot{H}}{H^{2}}\left( 2+\frac{\alpha }{H}%
\left( 1+\frac{8\pi G}{3}\rho _{C}\right) \right) +3\right].
\label{p}
\end{eqnarray}%
Eq. (\ref{p}) shows that $p_{C}\rightarrow -\rho _{C}$ whenever $\dot{H}%
\rightarrow 0$. It is also remarkable to note that, even without
assuming
Eq.~(\ref{emc1}), one can obtain Eq.~(\ref{e41}) and the second line of Eq.~(%
\ref{p}) by combining the time derivative of Eq.~(\ref{e4}) with
itself and using this fact that the $\rho $ source is
pressureless. In summary, we found out that the effects of
deviation from the Newtonian potential can be simulated as a
hypothetical fluid with energy density $\rho _{C}$ and pressure
$p_{C}$. The density parameter of this fluid can be calculated as

\begin{equation}
\Omega _{C}=\frac{8\pi G\rho _{C}}{3H^{2}}=\frac{\beta e^{-\frac{\alpha }{H}}%
}{1+\beta e^{-\frac{\alpha }{H}}}.  \label{dp}
\end{equation}

\noindent The deceleration and total state parameters of the model
can also be obtained by using $q_{C}=-1+(1+z)(dH(z)/dz)/H(z)$ and
$\omega _{C}\equiv p_{C}/\left( \rho _{C}+\rho \right) $,
respectively. Bearing the $\rho =\rho _{0}(1+z)^{3}$ in mind and
combining the above equation with the corresponding Friedmann
equation~(\ref{e4}), we easily reach

\begin{equation}
z(\Omega _{C})=\left( \frac{\gamma \left( 1-\Omega _{C}\right)
}{\ln
^{2}\left( \frac{\beta \left( 1-\Omega _{C}\right) }{\Omega _{C}}\right) }%
\right) ^{1/3}-1.  \label{zo1}
\end{equation}%
\noindent Eq.~(\ref{zo1}) implies that $z\geq -1$ for $\Omega
_{C}\leq 1$.
Calculations for the total state and deceleration parameters also lead to $%
\omega _{C}=\frac{2}{3}(q_{C}-\frac{1}{2})$ and

\begin{equation}
q_{C}=\frac{1-\Omega _{C}\ln \left( \frac{\Omega _{C}}{\beta (1-\Omega _{C})}%
\right) }{2+\Omega _{C}\ln \left( \frac{\Omega _{C}}{\beta (1-\Omega _{C})}%
\right) }.  \label{dpf2}
\end{equation}%
These results clearly show that $q_{C},\omega _{C}\rightarrow -1$
as $\Omega _{C}\rightarrow 1$ (or equally $z\rightarrow -1$) and
also $q_{C}\rightarrow \frac{1}{2}$ and $\omega _{C}\rightarrow 0$
as $\Omega _{C}\rightarrow 0$ (or equally $z\gg 1$).

In Fig.~(\ref{fig1}), we depict the behaviors of $\Omega _{C}$ and
$q_{C}$ with respect to $z$ by choosing suitable constants so that
the observation
constraints are satisfied. It is worthwhile to note again that the value of $%
\alpha $ can be found by specifying the value of $\rho _{0}$ from
observation and inserting it in $\alpha =\sqrt{8\pi \rho
_{0}G\gamma /3}$ (see below Eq.~(\ref{zo0})). In Fig.
(\ref{fig1}), we have plotted the behavior of $v_{sC}^{2}$ given
by

\begin{equation}
v_{sC}^{2}=\frac{dp_{C}}{d\rho _{C}}=\frac{dp_{C}/dH}{d\rho
_{C}/dH}. \label{vs}
\end{equation}%
Since the expression for $v_{sC}^{2}$ is too long, we have omitted
it here. As one can see from Fig. (\ref{fig1}), similar to type B
and many of the dark energy models \cite{stab,istab,istab1}, Type
C shows instability at present time too.

\section{Logarithmic Modification (type D) and Dark Energy\label{logsec}}

\begin{figure*}[t]
\includegraphics[width=.31\textwidth]{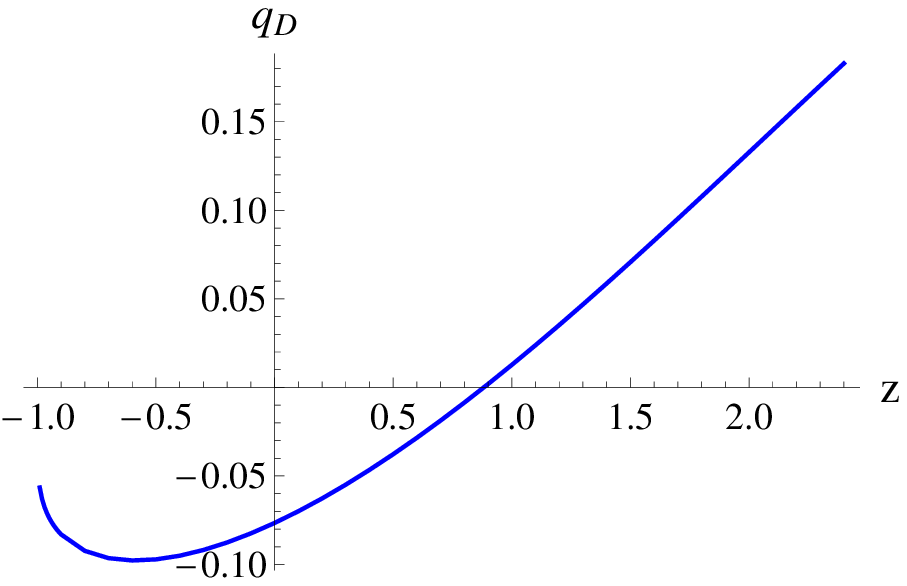}\label{phi1}
\includegraphics[width=.31\textwidth]{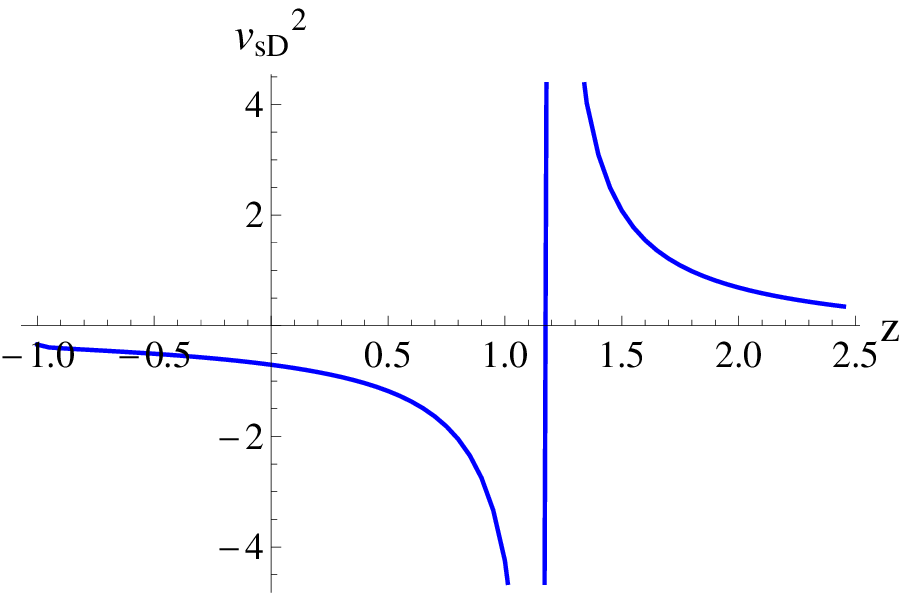}\label{stab2}
\includegraphics[width=.31\textwidth]{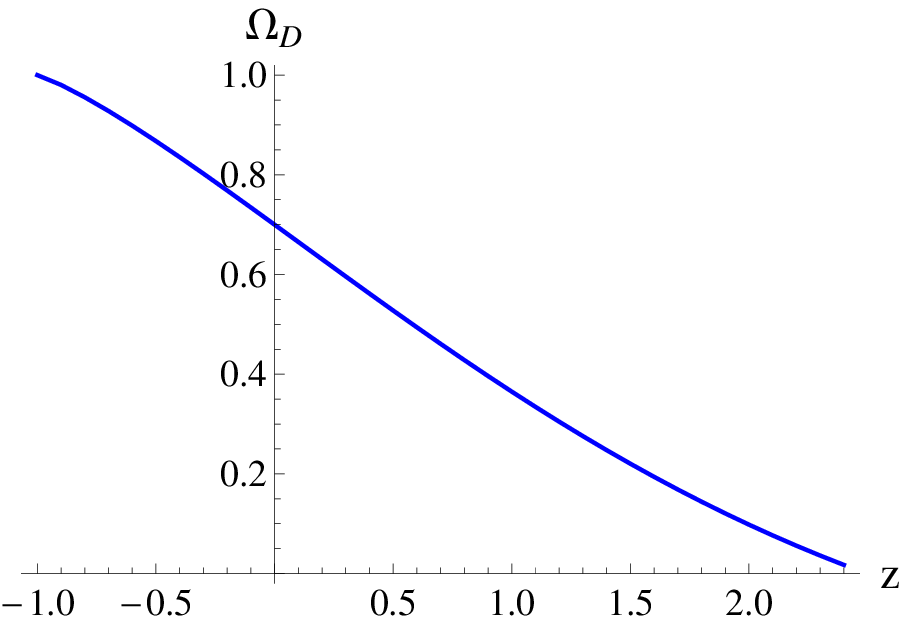}\label{phi2}
\caption{The behaviors of $\Omega _{D}$, $q_{D}$ and $v_{sD}^{2}$
in term of $z$ for $\protect\eta =240$ and $\protect\delta =122.5$
with $H\left( z=0\right) =67$ $km/\left( sMpc\right) $, $G=1$ and
the current matter density parameter $\Omega \left( z=0\right)
=0\cdot 3$.} \label{figphi}
\end{figure*}

Observations indicate that the universe is homogeneous and
isotropic in the scales larger than $100$-Mpc. Moreover, the
energy density of the dominant cosmic fluid ($\Lambda $) is
approximately constant at the mentioned scales \cite{roos}. In the
Newtonian language, these results can be summarized as

\begin{equation}
\frac{1}{r^{2}}\left[ \frac{d}{dr}\left( r^{2}\frac{d\phi }{dr}\right) %
\right] =4\pi G\Lambda ,  \label{ho1}
\end{equation}%
which finally leads to

\begin{equation}
\phi (r)=\frac{2\pi G}{3}\Lambda r^{2}-\frac{C}{r},  \label{ho2}
\end{equation}%
for the modified Kepler potential of the energy source confined to
the radius $r$. Here, $C$ is the integration constant, and in
order to cover the Kepler potential ($-GM/r$) at the appropriate
limit $\Lambda =0$, we should set $C=GM$, where $M$ is the mass
content of system. Thus, the modified Newtonian potential felt by
the test mass $m$ located at radius $r$ can be written as
\cite{mz,h1}

\begin{equation}
V(r)\equiv m\phi (r)=\frac{2\pi Gm}{3}\Lambda
r^{2}+V_{\mathrm{N}}(r). \label{ho30}
\end{equation}%
It means that if we modify the Newtonian potential by a Hook term
at the cosmic scales larger than $100$-Mpc, then the constant
energy density obtained by the observations may be justified. Such
potential can also be obtained in the non-local Newtonian gravity
framework \cite{ynl}. The light bending problem in the presence of
the above potential has also been studied in Ref. \cite{mz}. More
studies on the above potential can be found in Ref. \cite{h1} and
references therein.

Now, following the recipe used in this paper to find the Friedmann
equations in the Newtonian framework, one can easily reach

\begin{equation}
H^{2}=\frac{8\pi G}{3}\left( \rho +\frac{\Lambda }{2}\right) ,
\label{ho3}
\end{equation}%
which has an additional coefficient $1/2$ for the $\Lambda $ term
in comparison with the standard Friedmann equation in the presence
of the cosmological constant. This additional coefficient will be
disappeared if we modify the right hand side of Eq.~(\ref{ho1}) as
$8\pi G\Lambda $. It means that the flux corresponding to the
$\Lambda $ source is two times greater than those of the ordinary
known sources. In this manner, the modified Kepler
potential~(\ref{ho2}) finally takes the form

\begin{equation}
\phi (r)=\frac{4\pi G}{3}\Lambda r^{2}-\frac{GM}{r}.  \label{ho3f}
\end{equation}%
Also, the corresponding Friedmann equations will be the same as
those of the standard cosmology in the presence of the
cosmological constant.

The Kepler potential modified by a logarithmic term which can
describe the spiral galaxies rotation curves \cite{log1,log2,log3}
is written as

\begin{equation}
\Phi (r)=-\frac{GM}{r}-\delta GM\ln (\eta r),  \label{g1}
\end{equation}
where $\delta $ and $\eta $ are some constants found by fitting
the results with observations \cite{log1,log2,log3}. The potential
(\ref{g1}) can be obtained by various ways (see~\cite{ynl} and
references therein for details). Multiplying (\ref{g1}) by $m$,
one can get the corresponding Newtonian potential felt by the test
mass $m$ as

\begin{equation}
V(r)=V_{\mathrm{N}}(r)-\delta GmM\ln (\eta r),\text{ Type D}
\end{equation}

\noindent in accordance with the results of the non-local
Newtonian gravity \cite{ynl}. Calculations for the cosmological
equations corresponding to this potential lead to

\begin{equation}
3H^{2}=8\pi G\left( \rho +\rho _{D}\right) ,  \label{g3}
\end{equation}%
where the energy density of the pressureless source is $\rho =\rho
_{0}(1+z)^{3}$ and%
\begin{equation}
\rho _{D}=\frac{3H^{2}}{8\pi G}\left( \frac{\delta \ln (\frac{H}{\eta })}{%
\delta \ln (\frac{H}{\eta })-H}\right) .
\end{equation}%
One could also obtain%
\begin{equation*}
3H^{2}+2\dot{H}=-8\pi Gp_{D},
\end{equation*}%
where $p_{D}=-\dot{\rho}_{D}/3H-\rho _{D}$ is a pressure
originated from the logarithmic term. Hence, the density parameter
is $\Omega _{D}=\delta \ln
\left( \frac{H}{\eta }\right) /\left( \delta \ln \left( \frac{H}{\eta }%
\right) -H\right) $. Clearly, whenever $\delta >0$, we have
$\Omega
_{D}\rightarrow 0$ ($\Omega _{D}\rightarrow 1$) for $H\rightarrow \eta $ ($%
H\rightarrow 0$).

In Fig.~(\ref{figphi}), $\Omega _{D}$ and $q_{D}$ have been
plotted with
respect to $z$. We have not presented the explicit form of $%
q_{D}=-1+(1+z)(dH(z)/dz)/H(z)$ here since it is too long. Fig.
(\ref{figphi}) shows that $q_{D}\rightarrow \frac{1}{2}$ when
$z\rightarrow z_{\eta }$, where $z_{\eta}$ denotes the redshift at
which $H$ reaches its maximum possible value in this model, and
hence, the value of $H$ at the beginning of the matter dominated
era is a proper candidate for this maximum. Finally, we should
note that since the total state parameter of type D is defined as
$\omega_{D}\equiv p_{D}/\left( \rho +\rho_{D}\right)
=\frac{2}{3}(q_{D}-\frac{1}{2})$, it is obvious that $\omega_{D}$
also exhibits satisfactory behaviors.
$v_{sD}^{2}=dp_{D}/d\rho_{D}$ has been plotted in
Fig.~(\ref{figphi}). It is apparent that none of the obtained
models are stable for $z<0\cdot 1$.

\section{Concluding remarks}

Considering various modified Newtonian potentials, the Hubble law
and the classical total energy of a test mass located at the edge
of the universe, we could obtain some modified forms of the
Friedmann equations. In this formalism, it has been obtained that
some corrections to the Newtonian potential may model the current
accelerated universe, and hence, dark energy. The interesting
point is the ability of some of these corrected potentials in
describing the spiral galaxies rotation curves. The latter fact
signals that dark matter and dark energy may have some common
origins and aspects \cite{dara}. In fact, since all of the
obtained models address an accelerated universe, one may conclude
that the dark sides of cosmos may have at least some common roots.
Consequently, from this standpoint, a more complete modified
Newtonian potential may model the dark sides of cosmos
simultaneously.

\section*{Acknowledgments}
We are grateful to the anonymous reviewers for worthy comments.
MKZ would like to thank Shahid Chamran University of Ahvaz, Iran
for supporting this work. The work of H. Moradpour has been
supported financially by Research Institute for Astronomy \&
Astrophysics of Maragha (RIAAM) under research project No.
$1/5440-60$.

\end{document}